\documentclass[prd,superscriptaddress,twocolumn, 
preprintnumbers,showpacs,nofootinbib]{revtex4-1} 
\usepackage{amssymb,amsmath,graphicx,epsfig} 
\usepackage{hyperref,color,float,caption,array}
\usepackage{float}
\hypersetup{linktocpage=true}
\hypersetup{
colorlinks=true,   % false:boxed links; true:colored links
linkcolor=blue, % color of internal links
citecolor=blue, % color of links to bibliography
filecolor=blue,    % color of file links
urlcolor=blue      % color of external links
}
\def\beq{\begin{equation}} 
\def\eeq{\end{equation}} 
\def\bea{\begin{eqnarray}} 
\def\eea{\end{eqnarray}}

\begin{document} 
\title{Boosted di-boson from a mixed heavy stop}
%%%%%%%%%%%%%%%%%%%%%%%%%%%%%%%%%%%%%%%%%%%%%%%%%%%%%%%%%%%%%%%%%%%%% 
\author{Diptimoy Ghosh} 
\email{diptimoy.ghosh@roma1.infn.it} 
\affiliation{\normalfont{INFN, Sezione di Roma, 
Piazzale A. Moro 2, I-00185 Roma, Italy \\ and \\
Fermilab, P.O. Box 500, Batavia, IL 60510, USA}} 
%%%%%%%%%%%%%%%%%%%%%%%%%%%%%%%%%%%%%%%%%%%%%%%%%%%%%%%%%%%%%%%%%%%%% 
\begin{abstract} 

The lighter mass eigenstate ($\widetilde{t}_1$) of the two top 
squarks, the scalar superpartners of the top quark, is extremely 
difficult to discover if it is almost degenerate with the 
lightest neutralino ($\widetilde{\chi}_1^0$), the lightest and 
stable supersymmetric particle in the R-parity conserving 
supersymmetry. The current experimental bound on 
$\widetilde{t}_1$ mass in this scenario stands only around 200 
GeV. For such a light $\widetilde{t}_1$, the heavier top squark 
($\widetilde{t}_2$) can also be around the TeV scale. Moreover, 
the high value of the higgs ($h$) mass prefers the left and right 
handed top squarks to be highly mixed allowing the possibility of 
a considerable branching ratio for $\widetilde{t}_2 \to 
\widetilde{t}_1 h$ and $\widetilde{t}_2 \to \widetilde{t}_1 Z$. 
In this paper, we explore the above possibility together with the 
pair production of $\widetilde{t}_2$ $\widetilde{t}_2^*$ giving 
rise to the spectacular di-boson + missing transverse energy 
final state. For an approximately 1 TeV $\widetilde{t}_2$ and a 
few hundred GeV $\widetilde{t}_1$ the final state particles can 
be moderately boosted which encourages us to propose a novel 
search strategy employing the jet substructure technique to tag 
the boosted $h$ and $Z$. The reconstruction of the $h$ and $Z$ 
momenta also allows us to construct the stransverse mass $M_{T2}$ 
providing an additional efficient handle to fight the 
backgrounds. We show that a 4--5$\sigma$ signal can be observed 
at the 14 TeV LHC for $\sim$ 1 TeV $\widetilde{t}_2$ with 100 
fb$^{-1}$ integrated luminosity.

\end{abstract}
%%%%%%%%%%%%%%%%%%%%%%%%%%%%%%%%%%%%%%%%%%%%%%%%%%%%%%%%%%%%%%%%%
\preprint{FERMILAB-PUB-13-306-T}
\maketitle
%14.80.Da  :  Supersymmetric Higgs bosons 
%14.80.Ly  :  Supersymmetric partners of known particles  
%11.30.Pb  :  Supersymmetry
%%%%%%%%%%%%%%%%%%%%%%%%%%%%%%%%%%%%%%%%%%%%%%%%%%%%%%%%%%%%%%%%%
\section{Introduction}
%%%%%%%%%%%%%%%%%%%%%%%%%%%%%%%%%%%%%%%%%%%%%%%%%%%%%%%%%%%%%%%%%

A light third generation of superpartners remains an attractive 
possibility to realize weak scale Supersymmetry (SUSY) 
\cite{Martin:1997ns} in nature even after the successful 
completion of the 8 TeV run of the Large Hadron Collider (LHC). 
Moreover, if a light top squark (stop) $\widetilde{t}_1$ is the 
next-to-lightest SUSY particle (NLSP) just above the lightest 
neutralino ($\widetilde{\chi}_1^0$), the lightest and stable SUSY 
particle (LSP) in the R-parity conserving version of the 
Minimally Supersymmetric Standard Model (MSSM), it would have a 
significant density to coexist with the LSP around the freeze-out 
time, and annihilations involving $\widetilde{t}_1$ with the LSP 
\cite{Boehm:1999bj} can help achieve the LSP relic density 
consistent with the upper bound $\Omega^{\rm DM}h^2 < 0.128\,(3 
\sigma)$ presented by the Planck Collaboration 
\cite{Ade:2013zuv}.

Such a light $\widetilde{t}_1$ with a very small mass difference 
with the LSP ($\Delta m \equiv m_{\widetilde{t}_1} - 
m_{\widetilde{\chi}_1^0} \lesssim $ 50 GeV ) will dominantly 
decay to a charm quark and the LSP 
\cite{Hikasa:1987db,Muhlleitner:2011ww} resulting in a final 
state with jets and missing transverse \mbox{momentum 
($\slash{\hspace{-2mm}p}_T$)}. Owing to the small $\Delta m$, 
both the charm jet and the $\slash{\hspace{-2mm}p}_T$ will be 
extremely soft on average making this scenario very challenging 
to discover experimentally 
\cite{Carena:2008mj,He:2011tp,Drees:2012dd,Alves:2012ft, 
Agrawal:2013kha,Belanger:2013oka}.

The ATLAS collaboration has recently excluded a $\widetilde{t}_1$ 
mass of 200 GeV (95\% C.L.) in this channel for \mbox{$\Delta m < 
85$ GeV} using 20.3 fb$^{-1}$ of data collected at the 8 TeV run 
of the LHC \cite{ATLAS-CONF-2013-068}. This result clearly shows 
the low sensitivity of the current experimental searches to probe 
the degenerate stop NLSP region. Hence, it is extremely important 
to consider other possible signatures of light third generation 
SUSY in the stop NLSP scenario.

Interestingly, in this region of the SUSY parameter space the 
heavier stop $\widetilde{t}_2$ can also be below or around the 
TeV scale and it could prove useful to also look for them at the 
LHC. Motivated by this, we, in this paper, propose a novel search 
strategy to look for signatures of $\widetilde{t}_2$ at the 14 
TeV run of the LHC. Note that the signatures of $\widetilde{t}_2$ 
production at the LHC has not received enough attention in the 
literature in the recent past primarily because of the 
comparatively lower cross-section (due to its heaviness) 
\footnote{See however, \cite{Berenstein:2012fc} where the authors 
considered a very low $\widetilde{t}_2$ mass which in turn forced 
them to assume new F-term or D-term contributions to the higgs 
mass beyond the MSSM.} . However, with a few hundred $\rm 
fb^{-1}$ integrated luminosity expected at the 14 TeV LHC, the 
$\widetilde{t}_2$ production processes could be promising and can 
even provide information complimentary to the $\widetilde{t}_1$ 
production channels.

\begin{figure}[h!]
\begin{tabular}{c}
\includegraphics[width=0.8\columnwidth,height=0.6\columnwidth]
{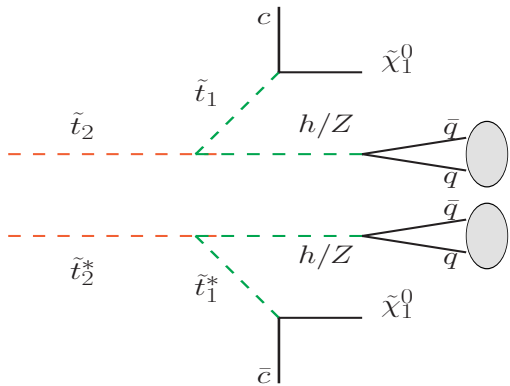}
\end{tabular}
\caption{Diagram showing a di-boson final state originating 
from $\widetilde{t}_2 \widetilde{t}_2^*$ pair production.
\label{fig1}}
\end{figure}

In this paper, we consider the pair production of 
$\widetilde{t}_2 \widetilde{t}_2^*$ and their subsequent decay to 
either $\widetilde{t}_1 Z$ or $\widetilde{t}_1 h$ final states 
(See Fig.\ref{fig1}). Note that, for the two above decays to have 
considerable branching ratios it is necessary to have adequate 
left-right mixing in the stop mass matrix 
\cite{Belanger:1999pv,Djouadi:1999dg}. Interestingly, the large 
radiative corrections to the higgs mass required for the consistency 
with the experimental observation also prefers the left and 
right handed top squarks to be highly mixed. Hence, the two decay 
modes $\widetilde{t}_2 \to \widetilde{t}_1 Z$ and 
$\widetilde{t}_2 \to \widetilde{t}_1 h$ are indeed very well 
motivated, in particular in the context of the higgs discovery. 
In addition, if most of the electroweak gauginos except the LSP, 
and the sbottoms are rather heavy which can indeed happen in a 
large region of the SUSY parameter space, the branching ratio 
of $\widetilde{t}_2$ to $\widetilde{t}_1 Z$ or $\widetilde{t}_1 h$ 
can be significant.

%%%%%%%%%%%%%%%%%%%%%%%%%%%%%%%%%%%%%%%%%%%%%%%%%%%%%%%%%%%%%%%%%
%%%%%%%%%%%%%%%%%%%%%%%%%%%%%%%%%%%%%%%%%%%%%%%%%%%%%%%%%%%%%%%%%

As we consider the stop NLSP scenario with a very small $\Delta 
m$, the $\widetilde{t}_1$ dominantly decays via 
$\widetilde{t}_1 \to c \, \widetilde{\chi}_1^0$. This gives rise 
to a final state consisting of $hh/hZ/ZZ$ + 
$\slash{\hspace{-2mm}p}_T$ + very soft jets. Moreover, for an 
approximately TeV scale $\widetilde{t}_2$ the decay products are 
sufficiently boosted and hence, a boosted di-boson system ($h$ or 
$Z$) along with moderately large $\slash{\hspace{-2mm}p}_T$ 
is the the experimental signature of such a scenario.

In the next section we will choose a couple of benchmark models 
where the specific decay chain mentioned above can be realized. 
The details of our event selection procedure will be discussed in 
sec.\ref{sig_bg}. In sec.\ref{results} we will present the final 
results for our signal as well as the backgrounds and stop 
thereafter with some concluding remarks.

%%%%%%%%%%%%%%%%%%%%%%%%%%%%%%%%%%%%%%%%%%%%%%%%%%%%%%%%%%%%%%%%%
\section{SUSY mass spectrum}
\label{spectrum}
%%%%%%%%%%%%%%%%%%%%%%%%%%%%%%%%%%%%%%%%%%%%%%%%%%%%%%%%%%%%%%%%%
 
\begin{figure}[h!]
\begin{tabular}{c}
\includegraphics[width=0.8\columnwidth,height=0.6\columnwidth]
{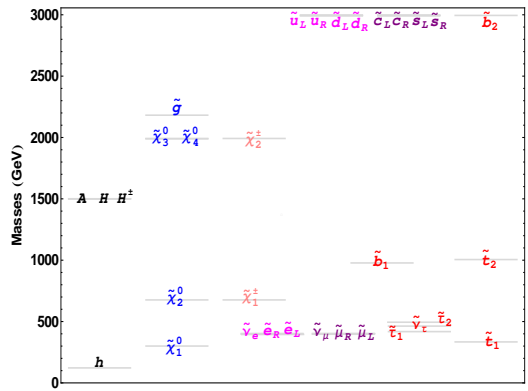}
\end{tabular}
\caption{Mass spectrum of the MSSM particles for our 
benchmark model:1. 
\label{fig2}}
\end{figure}

We now briefly discuss the MSSM mass spectrum relevant for our 
study and present a couple a benchmark models that will be used 
to present our results in sec.\ref{results}. In Fig.\ref{fig2} we 
graphically show the spectrum for one of our benchmark models 
(Model:1) where the two conditions
\begin{enumerate}
\item $\widetilde{t}_1$ is the NLSP with a small $\Delta m$,  
\item The branching ratios 
${\mathcal B}(\widetilde{t}_2 \to \widetilde{t}_1 \, Z)$ and/or 
${\mathcal B}(\widetilde{t}_2 \to \widetilde{t}_1 \, h)$ are 
significant, 
\end{enumerate}
can be realized. 

%%%%%%%%%%%%%%%%%%%%%%%%%%%%%%%%%%%%%%%%%%%%%%%%%%%%%%%%%%%%%%%%%
%%%%%%%%%%%%%%%%%%%%%%%%%%%%%%%%%%%%%%%%%%%%%%%%%%%%%%%%%%%%%%%%%
%
The most important input parameters of the phenomenological 
MSSM (pMSSM) which can be used to reproduce such a spectrum are 
shown in Table-\ref{tab:spectrum}.

Note that a low value of $\tan\beta$ and a high value of $m_A$ 
are motivated from the consistency of the measured branching 
ratios of the rare $B$-meson decays $B_s \to \mu^+ \mu^-$ and 
$B_d \to X_s \gamma$ with their SM predictions. In fact, we have 
checked that for both the benchmark models shown in 
Table-\ref{tab:spectrum} the SUSY predictions for the two above 
branching ratios are well inside the 2$\sigma$ experimental 
limits \cite{Asner:2010qj,Aaij:2013aka,Chatrchyan:2013bka}. 
The slepton masses 
are irrelevant for our discussion except the fact that the light 
sleptons can help ameliorate the discrepancy between the 
experimental measurement and the SM prediction of the anomalous 
magnetic moment of the muon 
\cite{Bennett:2006fi,Gnendiger:2013pva}. The relic density of the 
LSP for both the two benchmark models is also less than the 
Planck upper limit mentioned in the introduction. We have used 
the software package SuperIso Relic \cite{Arbey:2009gu} to 
calculate the $B$-decay branching ratios, anomalous magnetic 
moment of muon and the dark matter relic density.

%%%%%%%%%%%%%%%%%%%%%%%%%%%%%%%%%%%%%%%%%%%%%%%%%%%%%%%%%%%%%%%%%%%%%%
\begin{table}[h]
\centering
\begin{tabular}{|ccc|ccc|ccc|ccc|} 
\hline
\multicolumn{6}{|c|}{Model:1} & \multicolumn{6}{c|}{Model:2} \\
\hline
\multicolumn{3}{|c|}{pMSSM inputs} & \multicolumn{3}{c|}{Masses} & 
\multicolumn{3}{c|}{pMSSM inputs} & \multicolumn{3}{c|}{Masses}\\
\hline
$M_1$      &=&  300 & $m_{\widetilde{t}_2}$        &=& 1005 & 
$M_1$      &=&  410 & $m_{\widetilde{t}_2}$        &=& 1003 \\
$M_2$      &=&  650 & $m_{\widetilde{t}_1}$        &=&  334 & 
$M_2$      &=&  850 & $m_{\widetilde{t}_1}$        &=&  434 \\
$M_3$      &=& 2100 & $m_{\widetilde{\chi}_1^0}$   &=&  300 & 
$M_3$      &=& 2600 & $m_{\widetilde{\chi}_1^0}$   &=&  411 \\
$\mu$      &=& 2000 & $m_{\widetilde{\chi}_2^0}$   &=&  676 & 
$\mu$      &=& 2000 & $m_{\widetilde{\chi}_2^0}$   &=&  884 \\
$m_A$      &=& 1500 & $m_{\widetilde{\chi}_1^\pm}$ &=&  676 & 
$m_A$      &=& 1500 & $m_{\widetilde{\chi}_1^\pm}$ &=&  884 \\
$\tan\beta$&=&   10 & $m_h$                    &=&  123 & 
$\tan\beta$&=&  7.5 & $m_h$                    &=&  123 \\
$m_{Q3}$   &=& 1010 &                          & &      & 
$m_{Q3}$   &=& 1050 &                          & &      \\
$m_{t_R}$  &=&  630 &                          & &      & 
$m_{t_R}$  &=&  770 &                          & &      \\
$m_{b_R}$  &=& 3000 &                          & &      & 
$m_{b_R}$  &=& 3000 &                          & &      \\
$A_t$      &=&-1700 &                          & &      & 
$A_t$      &=&-1600 &                          & &      \\
\hline
\multicolumn{6}{|c|}
{${\mathcal B}(\widetilde{t}_2 \to \widetilde{t}_1 \, Z)$ = 52\%} & 
\multicolumn{6}{c|}
{${\mathcal B}(\widetilde{t}_2 \to \widetilde{t}_1 \, Z)$ = 56\%} \\  
\multicolumn{6}{|c|}
{${\mathcal B}(\widetilde{t}_2 \to \widetilde{t}_1 \, h)$ = 39\%} & 
\multicolumn{6}{c|}
{${\mathcal B}(\widetilde{t}_2 \to \widetilde{t}_1 \, h)$ = 41\%} \\
\multicolumn{6}{|c|}
{${\mathcal B}(\widetilde{t}_1 \to c \, \widetilde{\chi}_1^0)$ 
= 82\%}& 
\multicolumn{6}{c|}
{${\mathcal B}(\widetilde{t}_1 \to c \, \widetilde{\chi}_1^0)$ 
= 90\%}\\
\hline
\end{tabular}
\caption{The relevant pMSSM input parameters, particle masses 
and branching ratios for our two benchmark models. 
The mass spectrum and the branching ratios are calculated using the 
public package SUSY-HIT\cite{Djouadi:2006bz}. 
\label{tab:spectrum}}
\end{table}
%%%%%%%%%%%%%%%%%%%%%%%%%%%%%%%%%%%%%%%%%%%%%%%%%%%%%%%%%%%%%%%%%%%%

Note that the main difference between the two models in 
Table-\ref{tab:spectrum} is in the 
masses of $\widetilde{t}_1$ and $\widetilde{\chi}_1^0$. As the 
model:2 has higher values of these masses the mass gap between 
$\widetilde{t}_2$ and $\widetilde{t}_1$ is is smaller which in 
turn makes both the $\widetilde{t}_1$ and the $Z$ or $h$ less 
boosted.
 
%%%%%%%%%%%%%%%%%%%%%%%%%%%%%%%%%%%%%%%%%%%%%%%%%%%%%%%%%%%%%%%%%
\section{Collider Strategy}
\label{sig_bg}
%%%%%%%%%%%%%%%%%%%%%%%%%%%%%%%%%%%%%%%%%%%%%%%%%%%%%%%%%%%%%%%%%
As we already mentioned in the introduction, the relevant processes 
of our interest are 
%%%%%%%%%%%%%%%%%%%%%%%%%%%%%%%%%%%%%%%%%%%%%%%%%%%%%%%%%%%%%%%%%%%%%
$$
\begin{array}{ccccccc}
\rm p  p   &   \to   &  \rm \widetilde{t}_{2} \, \widetilde{t}_{2}^{*} &  \to 
&  \rm \widetilde{t}_{1} \, \rm \widetilde{t}_{1}^{*} \, Z \, Z & \to & 
Z \, Z \, \widetilde{\chi}_{1}^{0} \, \widetilde{\chi}_{1}^{0} \, c \, 
\bar{c} \\
&&&&&\hookrightarrow &  Z \, Z \, + \slash{\hspace{-2mm}p}_T + 
\textnormal{soft jets} \\
\rm p  p   &   \to   &  \rm \widetilde{t}_{2} \, \widetilde{t}_{2}^{*} &  \to 
&  \rm \widetilde{t}_{1} \, \rm \widetilde{t}_{1}^{*} \, Z \, h & \to & 
Z \, h \, \widetilde{\chi}_{1}^{0} \, \widetilde{\chi}_{1}^{0} \, c \, 
\bar{c} \\
&&&&&\hookrightarrow &  Z \, h \, + \slash{\hspace{-2mm}p}_T + 
\textnormal{soft jets} \\
\rm p  p   &   \to   &  \rm \widetilde{t}_{2} \, \widetilde{t}_{2}^{*} &  \to 
&  \rm \widetilde{t}_{1} \, \rm \widetilde{t}_{1}^{*} \, h \, h & \to & 
h \, h \, \widetilde{\chi}_{1}^{0} \, \widetilde{\chi}_{1}^{0} \, c \, 
\bar{c} \\
&&&&&\hookrightarrow &  h \, h \, + \slash{\hspace{-2mm}p}_T + 
\textnormal{soft jets} \, \, . \\
\end{array}
$$
%%%%%%%%%%%%%%%%%%%%%%%%%%%%%%%%%%%%%%%%%%%%%%%%%%%%%%%%%%%%%%%%%%%%%

There are several SM processes which can mimic our signal. They 
are all listed in Table-\ref{tab1}. Although, two $Z$ or higgs 
bosons are absent in most of the backgrounds, in practice there 
is always a possibility of $W$ boson being mis-tagged as a $Z$ or 
even a higgs boson. This fraction might not be very large but 
considering the gigantic cross sections for some of the 
backgrounds compared to the signal, the final contribution might 
not be negligible. Thus, a detailed simulation of all the 
background processes is necessary to make reliable predictions as 
we present in the next section.

As the mass gap between $\widetilde{t}_{2}$ and 
$\widetilde{t}_{1}$ is not so small in our case both the decay 
products of $\widetilde{t}_{2}$ namely, $\widetilde{t}_{1}$ and 
the $Z$ or the $h$ are expected to be moderately boosted. We thus 
consider the the fully hadronic decays of the $Z$ or the $h$ in 
order to be able to reconstruct them using the jet substructure 
technique.  Here we adopt the method proposed by Butterworth, 
Davison, Rubin and Salam (BDRS) \cite{Butterworth:2008iy} for 
tagging the hadronically decaying $Z$ or the higgs boson. We 
briefly describe below the exact procedure used in our analysis 
along with our other selection criteria.

As the first step of this algorithm, we construct ``fat-jets'' 
using the Cambridge-Aachen algorithm (CA algorithm) 
\cite{Dokshitzer:1997in} as implemented in the Fastjet package 
\cite{Cacciari:2005hq,Cacciari:2011ma} with $R$ parameter of 1.0. 
We demand that the fat-jets satisfy $p_{_{T}} > $ 200 GeV and 
pseudo rapidity $|\eta|$ $<$ 3.0. We then take a fat-jet $j$ and 
undo its last clustering step to get the two subjets $j_1$ and 
$j_2$ with $m_{j_1}$ $>$ $m_{j_2}$ by convention. The two 
quantities $\mu$= $m_{j_1}/m_{j}$ and $y = \Delta R^2_{j_1, j_2} 
\times {\rm min}(p^2_{T j_1}, p^2_{ T j_2})/m^2_{j}$, where 
$\Delta R_{j_1, j_2}$ is the distance between $j_1$ and $j_2$ in 
the $\eta$-$\phi$ plane, are then computed. If there is 
significant mass drop i.e., $\mu < \mu_c$ and the splitting of 
the fat-jet $j$ into $j_1$ and $j_2$ is fairly symmetric i.e., $y 
> y_c$ then we continue otherwise we redefine $ j = j_1$ and 
perform the same set of steps as describes above on $j_1$. The 
parameters $\mu_c$ and $y_c$ are tunable parameters of the 
algorithm and are set to 0.67 and 0.10 respectively in our 
analysis. If the above two conditions $\mu < \mu_c$ and $y > y_c$ 
are satisfied then the mother jet $j$ is taken and its 
constituents are re-clustered into CA jets with $R = R_{\rm filt} 
= {\rm min}(\Delta R_{j_1, j_2}/2, 0.4)$ resulting in a number of 
jets $j^{\rm filt}_1$, $j^{\rm filt}_2$, $j^{\rm filt}_3$, ..... 
$j^{\rm filt}_n$ ordered in descending $p_{T}$. The vector 
sum of the first three hardest jet momenta is then considered as 
the higgs candidate. The last step (the so called ``filtering'' 
procedure) is known to captures the dominant $\mathcal O 
(\alpha_{\rm s} )$ radiation from the higgs decay, while 
eliminating much of the contamination from underlying events 
\cite{Butterworth:2008iy}.

Once the BDRS procedure is applied on the fat-jets, we then 
impose the following selection criteria on the events:

%%%%%%%%%%%%%%%%%%%%%%%%%%%%%%%%%%%%%%%%%%%%%%%%%%%%%%%%%%%%%%%%%%%%%%
\begin{table*}[!t]
\begin{tabular}{|l|lr|c|p{10mm}|p{10mm}|p{10mm}|p{10mm}|p{10mm}|
p{10mm}|c|c|c|} 
\hline 
\multicolumn{4}{|}{} & 
\multicolumn{6}{|c|}{No. of events after } 
&\multicolumn{2}{c|}{}\\
\hline
Process & Production & & Simulated           
& S1 & S2 & S3 & S4 & S5 & S6 & Final & $\cal S $ \\
        &  cross-section &   & events &    
        &      &    &    &   &    &  cross-section (fb) 
        &  (100 fb$^{-1}$)\\
\hline 
\multicolumn{12}{|c|}{Signal} \\
\hline
Model:1 & 10 fb &\cite{Beenakker:1996ed}
& $10^{5}$ & 6012 & 4902 & 2736 & 2359 & 2143 & 1718 & 17.2 $\times 10^{-2}$ & 4.3\\ 
\hline
Model:2 & 10 fb &\cite{Beenakker:1996ed}
& $10^{5}$ & 6170 & 5319 & 2813 & 2421 & 2081 & 1853 & 18.5 $\times 10^{-2}$ & 4.6\\ 
\hline
\multicolumn{11}{|c|}{Backgrounds} \\
\cline{1-11}
$t \, \bar{t} $           & 833 pb &\cite{Aliev:2010zk}   
& $10^{8}$ &  221747 & 148580 & 142 & 41 & 26 & 11 & 9.1 $\times 10^{-2}$ \\ 
$t \, \bar{t} \, Z \, (1j) $      & 1.12 pb &\cite{Kardos:2011na}    
& 226110   &   2484  &  1444  &   8 &  7 &  1 &  1 & 0.5 $\times 10^{-2}$ \\ 
$t \, \bar{t} \, W^\pm \, (1j) $  & 770 fb   &\cite{Campbell:2012dh}  
& 276807   &   1365  &   787  &   5 &  3 &  3 &  2 & 0.5 $\times 10^{-2}$ \\ 
$t \, \bar{t} \, h \, (1j)$ & 700 fb   &\cite{Dawson:2003zu}    
& 231064   &   1893  &  1027  &   2 & 2  &  2 &  2 & 0.6 $\times 10^{-2}$  \\ 
$t / \bar{t} \; W^\pm \, (1j)$  & 64 pb &\cite{Campbell:2005bb} 
& 6518431 & 7596 & 5801 & 13 & 9 &  3 & 3 & 2.9 $\times 10^{-2}$\\ 
$P_1 \, P_2 \, P_3 (1j)$ & 500 fb  & \cite{Alwall:2011uj}
& 313350 & 1475 & 1093 & 10 & 5 & 4 & 2 & 0.3 $\times 10^{-2}$\\ 
$(P_i \in {W\,Z\,h})$  &     & 
&    &    &    &  &  &   &  &    \\ 
$P_1 \, P_2 + 1j/2 j$  & 5.5 pb  & \cite{Alwall:2011uj}
& 738779 & 2927 & 2646 & 3 & 3 & 3 & 3 & 2.2 $\times 10^{-2}$ \\ 
$(P_i \in {Z\,h})$  &     & 
&    &    &    &  &  &   &  &    \\ 
\cline{1-11}
Total & \multicolumn{9}{|c|}{} & 16.1 $\times 10^{-2}$ \\
Background & \multicolumn{9}{|c|}{} &  \\
\cline{1-11} 
\end{tabular}
\caption{Event summary after individual selection cuts both for 
the SUSY benchmark points as well as the SM backgrounds. See text 
for more details.
\label{tab1}} 
\end{table*} 

%%%%%%%%%%%%%%%%%%%%%%%%%%%%%%%%%%%%%%%%%%%%%%%%%%%%%%%%%%%%%%%%%%%%

\begin{itemize}

\item S1 : We demand that the two hardest fat-jets (with $p_T > 
200$\,GeV as mentioned before) reconstruct to either a $Z$ or a 
higgs boson with the mass windows [83.5-98.5] GeV and 
[118.5-133.5] GeV respectively.

\item S2 : As our signal events have no top quark in them but 
most of the backgrounds do, we find it useful to veto events 
which have top quark in them. In order to accomplish that we 
again construct fat-jets out of all the stable hadrons and apply 
the John Hopkins top tagger (JHToptagger) \cite{Kaplan:2008ie} 
on them. As the 
top quarks in the backgrounds are mostly not highly boosted we 
use a comparatively large $R$ parameter $R=1.6$ in order to not 
lose most of the top quarks from the backgrounds. We set the 
other parameters of the JHToptagger algorithm 
\cite{Kaplan:2008ie} to be $\delta_p=0.10$, $\rm \delta_r=0.19$, 
cos$\theta_{h}^{\rm max} = 0.7$ and the $W$ mass window = 
(60-100) GeV. We demand the final reconstructed top mass to fall 
in the window (158.3-188.3) GeV. We do not demand any b-tag for 
the reconstruction. An event is discarded if any of the 
fat-jets reconstructs to a top quark by the above criteria.

As we do not also expect any hard leptons in the signal we 
veto events which has any lepton with $p_T > 50$ GeV and 
$|\eta| < 2.5$. In addition to that, we also demand that the 
number of normal $R=0.4$ anti-$K_T$ jets \cite{Cacciari:2008gp} 
with $p_T > 50$ GeV and 
$|\eta| < 2.5$ be less than 6. Note that this step kills the 
multi-jet background configurations e.g., $t \bar{t}$ + additional 
hard jets while keeping almost all the signal events.

\item S3 : As $\widetilde{t}_2$ is quite heavier than both 
$\widetilde{t}_1$ and the $Z$ or $h$, the $\widetilde{t}_1$ in 
the decay $\widetilde{t}_2 \to \widetilde{t}_1 h/Z $ is expected 
to be rather energetic and a large part of its energy will be 
carried out by the LSP. Hence, even though $\Delta m$ is rather 
small the LSP will be quite energetic to record a high 
$\slash{\hspace{-2mm}p}_T$ in the detectors. A strong cut 
$\slash{\hspace{-2mm}p}_T > 400$ GeV reduces the backgrounds by a 
huge amount while keeping a handful of signal events.

\item S4 : The charm jet from the decay of $\widetilde{t}_1$ 
being very soft the topology of the decay $\widetilde{t}_2 \to 
\widetilde{t}_1 (\to c \widetilde{\chi}_1^0) Z/h $ looks exactly 
like the one where a mother particle decays to a visible daughter 
particle and an invisible particle. This observation motivates us 
to construct the stransverse mass $M_{T2}$ \cite{Lester:1999tx} 
out of the reconstructed $Z$ and/or the $h$ momenta and the missing 
transverse momentum. The $M_{T2}$ constructed in this way should 
be distributed till the $\widetilde{t}_2$ mass for the signal 
while the backgrounds are expected to populate the low mass 
region because there is no such heavy mother particle for the 
backgrounds. Requiring a large value of $M_{T2}$, $M_{T2} > 400$ 
GeV, helps us to tame the backgrounds efficiently.

\item S5 : The distribution of the effective mass of the system, 
$m_{\rm eff} = \Sigma \, p_{T}(\text{hard jets and hard leptons}) 
+ \slash{\hspace{-2mm}p}_T$, being strongly correlated to 
$2m_{\widetilde{t}_2}$ for the signal, is expected to occupy much 
higher mass region compared to the background. In our analysis, 
we require $m_{\rm eff} > 1250$ GeV to reduce the backgrounds 
further.

\item S6 : As the last step of our analysis we demand that there 
be no more than two normal jets (anti-$K_T$, $R=0.4$, $p_T > 50$ 
GeV and $|\eta| < 2.5$ as in S2) with $\Delta R > 1.0$ with the 
two reconstructed $Z$ or the $h$. We then also demand that none of 
these two jets are b-tagged. We use a 70\% efficiency for 
b-tagging and the rate for a c-jet (light jet) mis-tagged as a 
b-jet to be 15\% (1\%) \cite{CMS-PAS-BTV-09-001}. \end{itemize}

We use Pythia6.4.24\cite{Sjostrand:2006za} for generating the 
signal events. For most of the the backgrounds, we use Madgraph5 
\cite{Alwall:2011uj} to generate parton level events and 
subsequently use the Madgraph-Pythia6 interface (including 
matching of the matrix element hard partons and shower generated 
jets following the MLM prescription \cite{Hoche:2006ph} as 
implemented in Madgraph5) to 
perform the showering and implement our event selection cuts.

%%%%%%%%%%%%%%%%%%%%%%%%%%%%%%%%%%%%%%%%%%%%%%%%%%%%%%%%%%%%%%%%%
\section{Results and discussion}
\label{results}
%%%%%%%%%%%%%%%%%%%%%%%%%%%%%%%%%%%%%%%%%%%%%%%%%%%%%%%%%%%%%%%%%

In Table-\ref{tab1} we show the number of signal events for our two 
benchmark models as well as all the backgrounds after each of the 
selection criteria described in the previous section has been used. 
In the column:10 we show the final cross section when all the 
selection cuts have been imposed. 

Note that the number of events simulated for all the backgrounds 
is more than the numbers of events expected at the 14 TeV LHC 
with 100fb$^{-1}$ integrated luminosity. Thus, our estimation of 
the backgrounds is expected to be quite robust. In the first column 
the numbers within the brackets show the maximum number of additional jets 
which has been generated in Madgraph. For the $t \bar{t}$ background 
we have generated a huge number ($10^8$) of events in Pythia6 
(that means without additional hard jets) and checked that our 
background estimate agree with a smaller sample of MLM matched 
$t \bar{t}$ + jets events generated in Madgraph. We have also 
checked that the contribution of the processes 
$t/\bar{t} \; Z$ + jets, 
$t/\bar{t} \; h$ + jets, $W^+ W^-$ + jets, $W^\pm Z$ + jets 
and $W^\pm h$ + jets to the background is negligible.  

In the ultimate column of Table-\ref{tab1} we show the 
signal significance for an integrated luminosity 
of 100 fb$^{-1}$. While calculating the significance 
(${\mathcal S}$) we use the simple recipe of $S/\sqrt{B}$, 
$S$ and $B$ being the total number of signal and background events. 
Any additional systematic uncertainty (which is difficult to 
estimate in a phenomenological study) might change the 
significance somewhat but will not change the conclusion of 
our analysis in any significant way. It can be seen that 
that a significance ${\mathcal S} \sim$ 4-5 can be obtained 
with approximately 100 fb$^{-1}$ of data set.

In conclusion, we have considered the possibility of detecting a 
SUSY signal for the light stop NLSP scenario by considering the 
pair production of the heavier stop quark instead of the commonly 
considered light stop pair production channel. We have focused on 
the two decay channels $\widetilde{t}_2 \to \widetilde{t}_1 Z$ 
and $\widetilde{t}_2 \to \widetilde{t}_1 h$ giving rise to the 
spectacular di-boson + missing transverse energy final state. Employing the 
jet substructure techniques to reconstruct the hadronically decaying 
$Z$ and/or higgs 
momenta, which also enables us to construct the $M_{T2}$ out of 
the di-boson momenta and the $\slash{\hspace{-2mm}p}_T$, we have 
shown that a signal can be seen at the 14 TeV LHC with 
about 100 fb$^{-1}$ integrated luminosity. It is worth mentioning at this 
point that there is region of MSSM parameter space where the 
decay $\widetilde{t}_2 \to \widetilde{t}_1 h$ itself can also be 
substantial \cite{Belanger:1999pv} and the possibility 
to see even just the di-higgs signal should 
be investigated. This could also be important in the context of 
higgs self-coupling measurements which is extremely important in our endeavor 
to look for signatures of new physics beyond the SM.

\vspace{2mm}
%%%%%%%%%%%%%%%%%%%%%%%%%%%%%%%%%%%%%%%%%%%%%%%%%%%%%%%%%%%%%%%%%%
%\clearpage
\acknowledgments
%%%%%%%%%%%%%%%%%%%%%%%%%%%%%%%%%%%%%%%%%%%%%%%%%%%%%%%%%%%%%%%%%%
We thank JoAnne Hewett and Tom Rizzo for the hospitality at the 
SLAC Theory Group where this work was started. 
We also thank Prateek Agrawal, Wolfgang Altmannshofer, Martin Bauer, 
Patrick Fox, Raoul R\"ontsch and Felix Yu for useful discussions. 
The research leading to these results has received funding from 
the European Research Council under the European Union's Seventh 
Framework Programme (FP/2007-2013) / ERC Grant Agreement 
n.279972.

%%%%%%%%%%%%%%%%%%%%%%%%%%%%%%%%%%%%%%%%%%%%%%%%%%%%%%%%%%%%%%%%%%%
%\bibliographystyle{h-physrev5}
%\bibliography{st2st1h}

%%%%%%%%%%%%%%%%%%%%%%%%%%%%%%%%%%%%%%%%%%%%%%%%%%%%%%%%%%%%%%%%%
\end{document}